\documentclass[12pt]{article}

\usepackage{amsmath,amssymb}
\usepackage{setspace,color,enumitem,url,soul,arydshln,float}

\def\bs{\boldsymbol}
\def\mc{\mathcal}

\usepackage{authblk}
\author{\vspace{-3mm} Li-Chun Zhang}
\affil{\em \vspace{-5mm} University of Southampton}
\title{\LARGE \vspace{-15mm} On provision of UK neighbourhood population statistics beyond 2021}
\date{}

\begin{document}

\maketitle

\paragraph{Summary} Census 2021 may well be the last of its kind in the UK. For provision of population statistics in the immediate years following 2021, the basic scheme currently envisaged is to supplement available administrate data with a continuous coverage survey, which amounts to a yearly sample size of about 0.5 million addresses, although the details of the methodology are yet to be determined. Meanwhile, the ONS is seeking alternative approaches, which can make greater use of the relevant administrative data. This report outlines the basic ideas of a \emph{rolling approach} for provision of  \emph{UK neighbourhood population statistics} beyond 2021, set in the broad perspective of establishing a sustainable future for official statistical systems, which is faster in response, richer in detail and greater in return of long-term cost efficiency (Zhang, 2013). It consists of 4 sections:
\begin{description}
\item{1.} Selective review of post-census method

\item{2.} Fractional counting and rolling

\item{3.} Transition to register-based auditing-assisted approach

\item{4.} Key topics for methodology development  
\end{description}

\section{Selective review of post-census method}

Internationally and historically speaking, one can distinguish three main approaches when it comes to the provision of post-census population statistics: Demographic Balancing Equation, Central Population Register and Adjusted Population Dataset. Some selected examples of each approach are briefly reviewed below.

\subsection{Demographic Balancing Equation} 

The Demographic Balancing Equation (DBE) allows one to update census population statistics, based on vital statistics, available data on internal and external migration, and various group-quarter (GQ) or other special populations (e.g. military personals). 
\begin{itemize}[leftmargin=4mm]
\item In reality the approach is based on administrative data of birth, death, internal migration and GQ or special populations. Nevertheless, for various reasons, longitudinal linkage of relevant data and the base (census) population at the individual level is not the case in practice. The basis of production is a yearly constructed population hypercube rather than, say, a one-number census-like population dataset.

\item Survey data are often required for the external migration component. Indirect residual estimation methods are sometimes used for certain migration groups and special populations. In the UK context, it may be possible to make greater use of the continuous coverage survey under the DBE approach, as will be commented later.  
\end{itemize}

\subsubsection{Example of USA}

In the USA, the Population Estimates Program (PEP) uses the DBE approach, supplemented by the American Community Survey (ACS) for foreign-born immigration.

The PEP produces population estimates at three levels: national, state and county, by characteristics of sex, age, race and Hispanic origin; see e.g. CBS (2019). At the first stage, the national characteristics, state total, and county total estimates are created. At the second stage, estimates of states and counties by characteristics are obtained, by raking of the lower-level DBE-estimates to the controlled estimates at a higher level. Linear interpolation between two successive yearly estimates generates the seemingly continuous Population Clock, which can be viewed at \url{http://www.census.gov/popclock/}. 

The DBE-estimates are essentially based on administrative data, except for foreign-born immigration, which uses the ACS. In particular, for state and county total estimates, one calculates county-to-county net domestic migration based on four sources: Internal Revenue Service tax return data for ages 0-64, Medicare enrollment data from Centers of Medicare and Medicaid Services for ages 65+, Social Security Administration's Numerical Identification File for all ages, and change in the GQ population.

The ACS uses an address frame, and aims primarily at producing attribute statistics similar to the census long-form format. Two ACS weights are computed for households and persons, respectively. The household unit weight is produced first, based on sampling design inclusion probabilities reweighted for nonresponse and then raked to county/group of county level totals; person weight is then produced by raking. See e.g. CBS (2014).

\subsubsection{Comments on UK}

The DBE approach is used to produce the mid year population estimates (MYE) in the UK; see e.g. ONS (2018). The key dissemination levels are national and local authority. Internal migration data are derived from administrative sources. For international migration one makes use of the International Passenger Survey, while administrative sources are used to distribute immigrants to each local authority.

One anticipates the relevant administrative sources to be enhanced in the coming years. The most relevant sources include the Benefits and Income data, the Patient Register, Education data at all levels, the Council Tax data, and so on. In particular, updates of home addresses in the Drivers' License database may potentially prove to be an important addition, regarding internal migration. Moreover, Migrant Work Scan data can provide information on all overseas nationals who have registered for and allocated a National Insurance Number. Together with Exit Checks they could be expected to greatly improve the external migration data in future.

Going forward, two important questions need to be considered for potential improvements of the estimation of DBE components in the UK context. 
\begin{itemize}[leftmargin=4mm]
\item Can the ability to link individual-level administrative data across sources improve the quality of address or locality information, and extend the range of sources to be used? For instance, in the USA, one splits the non-GQ population into 0-64 and 65+ by age, each with its designated source, which may be simplistic and practical, only if individual linkage is infeasible but not otherwise.

\item Can greater use be made of the coverage survey in the DBE approach in the UK, compared to the ACS in the USA? The question is natural, especially provided linkage to the administrative data. But even without linkage, it may be possible to combine area-level administrative and survey migration data, e.g. by means of generalised structure preserving estimation for small area compositions.
\end{itemize}

\subsection{Central Population Register}

In a number of countries that produce register-based census-like population statistics, Central Population Register (CPR) is used for continuous updating of neighbourhood population statistics at very low aggregation levels. A wide range of statistics can be produced with much greater ease and lower cost, including timely population dynamics, detailed migration flows and household statistics that can inform policy makers, researchers, businesses and general public. See e.g. some statistics produced by Statistics Norway at \url{https://www.ssb.no/befolkning/faktaside/befolkningen#blokk-1}.

There are two key political and cultural premises of the CPR approach: infrastructure and population concept. The CPR approach is only feasible, given the necessary legal framework, uptake of universal person identification in public services, and adequate time labelling of relevant demographic events and recording in the CPR. It requires also the population to be counted at their \emph{de jure} instead of \emph{de facto} address. The latter poses a key challenge to the relevance of the statistics, in which respect statistical adjustment may still be necessary for certain topics. See e.g. Zhang (2011) and Zhang and Fosen (2018) for a discussion of register-based household statistics.

\subsection{Adjusted Population Dataset}

In some countries, the CPR does not have the desired accuracy to warrant  the CPR approach directly, often due to a lack of updated migration data. Moreover, in countries where CPR does not exist at all, it may still be possible to construct a statistical Population Dataset (PD), based on linkage and integration of relevant administrative registers, which can yield national population counts that are similar to the census population estimates, as it is e.g. the case in the UK (ONS, 2015), New Zealand and Australia. In either case, statistical adjustment is then necessary. Despite nearly all the examples below are based on adjustment of the CPR, we shall refer to this approach as \textit{Adjusted Population Dataset (APD)}, in anticipation of future developments that can enable the approach in those countries that do not have CPR at all. 

The main issues for adjustment are \emph{erroneous enumeration}, \emph{missing enumeration} and \emph{misplacement} in the PD. Erroneous and missing enumerations causes over- and under-counting at the national level, respectively; whereas misplacement causes inter-locality over- and under-counting simultaneously but does not cause over- or under-counting at the national level. Below we give some examples of the different cases.

\subsubsection{APD for misplacement in Israel}

The CPR in Israel has only negligible over- and under-counting errors at the national level. The Israeli Integrated Census 2008 was conducted chiefly to adjust for misplacement in the CPR. See e.g. Nirel and Glickman (2009). All the persons registered in a Statistical Area (SA) according to the CPR were given a weight, such that their weighted total equals to the estimated population size in that SA. There are over 3000 SAs in Israel.

For post-census population statistics, a person's weight would remain the same, as long as the person stays in the same SA. A person would in principle take on the weight of the destination SA, if the person registers a move to another SA in the CPR. In this way, the pattern of misplacement adjustment in each SA is preserved, while the CPR count of persons registered in the SA naturally varies over time.

Notice that in reality the basic idea expounded above needs to be refined in several ways. For instance, the weight may vary across different sub-populations within an SA. Moreover, ad hoc adjustment may be required, if one detects an abnormal change in a given SA, often due to delays of previous or new housing developments.

\subsubsection{APD for erroneous enumeration in Latvia and Estonia}

Erroneous enumerations due to lack of updating of emigration were evidenced in the last census in both Latvia and Estonia. Each developed new method for post-census population statistics, which are similar in some respects while differ in others.

The initial census 2011 enumeration returned a population count of 2.075 million in Latvia, which was about 7\% lower than the CPR count. The Central Statistical Bureau of Latvia worked out an adjustment method based on statistical classification and migration mirror statistics (CSBL, 2019). By combining the census data with relevant administrative data including the CPR, approximately 100 000 persons were added to the census enumeration from the administrative data. Treating the imputed census enumeration as the labelled units, supervised learning by logistic regression yielded a predicted probability of erroneous enumeration for each eligible person in the CPR. The fitted model was used in the post-census years to adjust the CPR population count. 

The imputation of census under-enumeration was similar in Estonia, which added about 30 000 persons from the relevant administrative sources to the initial census enumeration of about 1.3 million in total. A residency index was developed for the post-census years, which is updated on a yearly basis (Tiit and Maasing, 2016). First, an extended population $(U_+)$ is constructed, which has negligible under-coverage errors. Next, 27 administrative sources are used to construct a Sign-of-Life score $X(k,t-1)$ for person $k\in U_+$ in year $t$, including special care, parental leave, dental care, digital prescription, prison visit, change of vehicle, residence permit, and so on. The residency index for person $k$ in year $t$, denoted by $R(k,t)$ is calculated from $R(k,t-1)$ and $X(k,t-1)$ as
\[
R(k, t) = d\cdot R(k, t-1) + g\cdot X(k, t-1)
\]
where $d$ is the stability rate and $g$ the signs of life rate, the values of which are heuristically chosen to yield plausible updated population counts.

\subsubsection{APD for erroneous and missing enumerations in Italy}

The population census in Italy is moving to a `permanent' census, which will produce annual population statistics instead of the  previous decennial cycle, using information from administrative sources integrated with sample surveys. Moreover, the first-phase sample for population statistics will provide the frame for the main social survey samples, which are negatively coordinated at the second phase.

The first-phase sample has two components. The component A consist of a sample of Enumeration Areas or addresses selected from an Integrated Address File. The component L is selected from a list of households (in the CPR), to provide reliable information on the `census' variables that are not available from the administrative sources. The two components A and L will amount to a yearly sample size of about 400,000 households and 1,000,000 persons, respectively, drawn from 2850 out of 7950 municipalities.

Population estimates will be obtained by an extended Dual System Estimation method, accounting for both under-coverage and over-coverage errors of the CPR. It is planned that all the eligible persons in the CPR will be given a weight, such that the weighted totals are equal to the estimated population sizes in the respective municipalities.

\subsubsection{APD for missing enumerations in Ireland}

The traditional population census in Ireland takes place every 5 years. Despite the absence of CPR, the CEO has developed an alternative estimation methodology at the national level, based purely on administrative sources; see e.g. Dunne (2015), Zhang and Dunne (2017). The Irish APD approach is unique internationally, where it operates in such a way that one only needs to deal with the missing enumerations at the national level.

The core of Dual System Estimation consists of two lists both subjected to missing enumerations only. The CEO constructs a Sign-of-Life register, called the Person Activity Register (PAR), based on observed activities across a range of administrative sources, such that the PAR is expected to have only negligible over-counting errors, whereas it can have systematic under-coverage errors in different parts of the usual resident population. Next, the Drivers' License Database (DLD) provides a plausible second cross-sectional enumeration list, based on the fact that each holder needs to renew the license every 10 years. As discussed in Zhang (2019b), the two lists PAR and DLD can satisfy the assumptions of Dual System Estimation, which differ from those assumptions traditionally held for estimation based on census and census coverage surveys (Wolter, 1986).

\section{Fractional counting and rolling}

The CPR approach is unlikely to be feasible in the UK in the near future beyond 2021. Provided linkage across the sources, the APD approach will encompass the DBE approach, now that all the administrative data for component estimation can be brought together into an integrated PD. The key challenge then, for making greater use of the combined administrative and survey data, will be to achieve an individual-based estimation methodology (such as in Israel, Latvia and Estonia), instead of population hypercube-based estimation (such as currently envisaged in Italy and Ireland). This requires above all two extensions to the existing APD approach: 
\begin{itemize}[leftmargin=6mm]
\item whereas the current Israeli approach focuses only on misplacement and the Latvian and Estonian approaches only on erroneous enumeration, one needs to be able to account for more than one type of error in the APD approach for the UK;

\item one needs a more rigorous methodology for the updating of weights, score or index over time, for the individuals in the PD. 
\end{itemize}
Below, to address the first issue, we outline a theory of \emph{fractional counting} for population statistics; regarding the second issue, we discuss the basic ideas for the rolling or incremental learning of the fractional counters.

\subsection{Fractional counting for misplacement}

To focus on the idea, suppose for the moment that the PD, denoted by $\mc{P}$, is only subjected to misplacement, where $\mc{P} = U$ and $U$ is the target population.

\paragraph{Sign-of-Life (SoL) addresses} For each person $k$ in $\mc{P}$, one finds the possible distinct addresses, at which the person may be located according to all available administrative sources. For instance, a student may have home address (of the parents) in addition to an address in the Higher Education loan register. Or a person may have different addresses in the Patient Register and the Council Tax register.  Notice that depending on the data available and the aggregation detail required for population statistics, the address may identify a coarser-than-dwelling location, such as post code or municipality. We shall refer to such addresses as the SoL-addresses.

\paragraph{SoL-address classifier and predictor} Let ${\bf a}_k$ be the $q$-vector containing all the available SoL-addresses of person $k$ in $\mc{P}$. Let ${\bf z}_k$ contain all the relevant auxiliary data, such as known family relationships, emigration status, previous addresses, work or study place, and so on. A SoL-address \textit{classifier} is given by
\[
{\bf y}_k = g({\bf a}_k, {\bf z}_k)  \in \{ 0, 1\}^q\qquad\text{where}\quad {\bf y}_k^{\top} {\bf 1} = 1 
\]
i.e. one and only one of the available addresses is chosen as the address for person $k$, such that the corresponding component of ${\bf y}_k$ is set to 1 and all the other components are set to 0. Moreover, a SoL-address \textit{predictor}, or \emph{fractional counter}, is given by
\[
\bs{\mu}_k = h({\bf a}_k, {\bf z}_k)  \in [0, 1]^q\qquad\text{where}\quad \bs{\mu}_k^{\top} {\bf 1} = 1  
\]
i.e. each component can take value from 0 to 1 and all the components sum to 1. The idea is for each component of $\bs{\mu}_k$ to be probability that the corresponding address is the true usual resident address of person $k$, denoted by adr$_k$.

\paragraph{Population size based on fractional counting} Based on the individual classifier for all $k\in \mc{P}$, the population count of locality $i$, for $i = 1, ..., m$, is given by
\begin{equation}\label{cls}
\widehat{N}_i^C = \sum_{k\in \mc{P}} {\bf y}_k^{\top} \bs{\delta}_k \qquad\text{and}\qquad \bs{\delta}_k = \bs{\delta}({\bf a}_k \in A_i) 
\end{equation}
where $A_i$ denotes the set of admissible addresses for the $i$th locality, and $\bs{\delta}({\bf a}_k \in A_i)$ is the $q$-vector with each component taking value 1 if the corresponding address belongs to $A_i$ and 0 otherwise. Whereas, based on the fractional counter $\bs{\mu}_k$ and the method of \emph{fractional counting}, the population count of the same locality is given by
\begin{equation}\label{pred}
\widehat{N}_i^P = \sum_{k\in \mc{P}} \bs{\mu}_k^{\top} \bs{\delta}_k \qquad\text{and}\qquad \bs{\delta}_k = \bs{\delta}({\bf a}_k \in A_i) 
\end{equation}

\paragraph{Properties of fractional counting} Prediction by classifier \eqref{cls} resembles election by majority vote, where the winner takes \emph{all} the votes regardless of the margin over the votes for the loser. It will cause bias of population statistics. Prediction by fractional counting \eqref{pred} aims to avoid this problem. It is unbiased for any $N_i$, provided
\begin{equation} \label{unbias}
\begin{cases} \bs{1}^{\top} \mbox{Pr}(\mbox{adr}_k = {\bf a}_k) = 1 \\ 
\bs{\mu}_k = \bs{\mu}({\bf a}_k, {\bf z}_k) \end{cases} 
\end{equation}
The first condition ensures that the true address (adr$_{k}$) can only be one of the SoL-addresses, insofar misplacement is the only problem at hand. The second condition then ensures that the probabilities of $\mbox{adr}_k = \bs{a}_k$ is entirely determined by ${\bf a}_k$ and $\bs{z}_k$. In reality the matter will depend on the covariates $\bs{z}_k$ and how well $\bs{\mu}_k$ in \eqref{pred} is modelled. Given the $\bs{\mu}_k$'s, the prediction variance of $\widehat{N}_i^P$ by \eqref{pred} is 
\[
V(\widehat{N}_i^P - N_i) = \sum_{k\in \mc{P}} \bs{\mu}_k^{\top}  \bs{\delta}_k \big( 1 - \bs{\mu}_k^{\top}  \bs{\delta}_k \big)
\]
where we assume that $\delta(\mbox{adr}_k \in A_i)$ is independent across different persons, conditional on the $({\bf a}_k, {\bf z}_k)$'s. However, it is possible to allow for clustering effects in the variance calculation, depending on the model underlying $\bs{\mu}_k$, such as when it always assigns the same $\bs{\mu}_k$ to all the persons in the same family and family relationship is part of $\bs{z}_k$.

\paragraph{Producing social statistics} Population and sub-population totals based on fractional counting can provide calibration totals for social surveys in the same way as the MYEs. Consider register-based attribute statistics. Denote by $\epsilon_k$ the value of interest for $k\in \mc{P}$, and $\hat{\epsilon}_k$  the corresponding register-based value. In cases where $\epsilon_k$ is observed without error, one can simply set $V(\hat{\epsilon}_k - \epsilon_k) = 0$; otherwise the variance will be positive in cases of model-based prediction of $\epsilon_k$ based on register sources. The population and fractional counting totals in the $i$th locality are given, respectively, as
\[
t_i = \sum_{k\in \mc{P}} \delta_k \epsilon_k \qquad \text{and}\qquad \hat{t}_i = \sum_{k\in \mc{P}} \hat{\mu}_k \hat{\epsilon}_k
\]
where $\delta_k = \delta(\mbox{adr}_k \in A_i)$ as defined in \eqref{cls} and \eqref{pred}, and $\hat{\mu}_k = \bs{\mu}_k^{\top} \bs{\delta}_k = \widehat{\mbox{Pr}}(k \in U_i)$ by the fractional counter. We have unbiased $\hat{t}_i$, for $i = 1, ,..., m$, provided
\[
E(\hat{\mu}_k - \delta_k) = 0 \quad\text{and}\quad E(\hat{\epsilon}_k - \epsilon_k) =0 \quad\text{and}\quad
(\hat{\epsilon}_k, \epsilon_k) \perp (\hat{\mu}_k, \delta_k)~, 
\]
since $E(\hat{t}_i - t_i) = \sum_{k\in \mc{P}} E(\hat{\mu}_k \hat{\epsilon}_k - \delta_k \epsilon_k)$, where 
\[
E(\hat{\mu}_k \hat{\epsilon}_k - \delta_k \epsilon_k) = 
E(\hat{\mu}_k \hat{\epsilon}_k - \delta_k \hat{\epsilon}_k + \delta_k \hat{\epsilon}_k - \delta_k \epsilon_k) = 0 ~.
\]
Provided independence across different persons, the prediction variance is given as
\[
V(\hat{t}_i - t_i) = \sum_{k\in \mc{P}} V(\hat{t}_i - t_i)
\]
Again, it is possible to relax the independence assumption and allow for clustering effects in the variance calculation.

\subsection{Initiation in the absence of missing enumerations}

It is natural to initiate the fractional counters in connection with the next census. The pre-2011 MYEs were retrospectively adjusted upwards given the census population estimates. This suggests that the administrative sources for the DBE component estimation may suffer from some under-coverage. Nevertheless, we shall assume that going forward the enhanced administrative sources will enable one to construct the PD as an extended population, denoted by $\mc{P}$, which has negligible under-coverage of the population, denoted by $U$. However, for any in-scope person, we do not require the first condition of \eqref{unbias} to hold, in anticipation of possible weakness of the SoL-address sources. That is, for each person $k\in \mc{P} \cap U$, we allow for a probability of being \emph{displaced}, denoted by
\[
\xi_k = 1 - \bs{1}^{\top} \mbox{Pr}(\mbox{adr}_k = {\bf a}_k)  \geq 0 ~.
\]

For supervised learning of \emph{fractional counter}, one ideally needs to label everyone in $\mc{P}$ first as erroneous or not, then in the latter case, displaced or not, and finally in the case of not displaced, the vector $\mbox{adr}_k = {\bf a}_k$. The situation following the UK census 2021 is ragged due to the presence of multiple errors, as depicted in Table \ref{tab-initial}.

\begin{table}[ht]
\begin{center}
\caption{Initiation of fractional counters in $\mc{P}$ based on census}
\begin{tabular}{cccccccccccc} \hline
\multicolumn{8}{c}{Fractional counter} & & & & Enum. \\ \cline{1-8}
\multicolumn{4}{c}{Placement} & $~$ & Displaced & $~$ & Erroneous & $~$ & $\mc{P}$ & $~$ & Census \\ 
\cline{4-4} \cline{6-6} \cline{8-8} \cline{10-10} \cline{12-12}
& & & \multicolumn{1}{|r|}{1 adr} & & \multicolumn{1}{|c|}{$\xi_1$} & & \multicolumn{1}{|c|}{$\theta_1$} & 
& \multicolumn{1}{|c|}{1} & &  \multicolumn{1}{|c|}{1} \\ \cline{3-4}
& & \multicolumn{2}{|r|}{2 adr} & & \multicolumn{1}{|c|}{$\vdots$} & & \multicolumn{1}{|c|}{$\vdots$} & 
& \multicolumn{1}{|c|}{$\vdots$} & Core & \multicolumn{1}{|c|}{$\vdots$} \\ \cline{2-4}
& \multicolumn{3}{|r|}{3 adr} & & \multicolumn{1}{|c|}{$\vdots$} & & \multicolumn{1}{|c|}{$\vdots$} & 
& \multicolumn{1}{|c|}{$\vdots$} & &  \multicolumn{1}{|c|}{$\vdots$} \\ \cline{1-4}
\multicolumn{4}{|r|}{$\cdots\quad\cdots\quad\vdots$} & & \multicolumn{1}{|c|}{$\xi_{N_c}$} & & \multicolumn{1}{|c|}{$\theta_{N_c}$} & 
& \multicolumn{1}{|c|}{$N_c$} & &  \multicolumn{1}{|c|}{$N_c$} \\ \cline{1-4} \cdashline{5-12}
& & & \multicolumn{1}{|r|}{1 adr} & & \multicolumn{1}{|c|}{$\vdots$} & & \multicolumn{1}{|c|}{$\vdots$} &
& \multicolumn{1}{|c|}{$\vdots$} & & \multicolumn{1}{|c|}{$\vdots$} \\ \cline{3-4}
& & \multicolumn{2}{|r|}{2 adr} & & \multicolumn{1}{|c|}{$\vdots$} & & \multicolumn{1}{|c|}{$\vdots$} &  
& \multicolumn{1}{|c|}{$\vdots$} & (Non- & \multicolumn{1}{|c|}{$N_L$} \\ \cline{2-4}  \cline{12-12}
& \multicolumn{3}{|r|}{3 adr} & & \multicolumn{1}{|c|}{$\vdots$} & & \multicolumn{1}{|c|}{$\vdots$} & 
& \multicolumn{1}{|c|}{$\vdots$} & core) & $\widehat{N}$ \\ \cline{1-4} 
\multicolumn{4}{|r|}{$\cdots\quad\cdots\quad\vdots$} & & \multicolumn{1}{|c|}{$\xi_{N_p}$} & & \multicolumn{1}{|c|}{$\theta_{N_p}$} & & \multicolumn{1}{|c|}{$N_p$}  \\ \cline{1-4} \cline{6-6} \cline{8-8} \cline{10-10} \\
\multicolumn{8}{c}{$\underbrace{\sum \bs{\mu}_k^{\top} \bs{1} + \sum \xi_k = \widehat{N}, \quad \widehat{N} + \sum \theta_k = N_p}_\text{Benchmarking}$}
\end{tabular} \label{tab-initial}
\end{center}\end{table}

Reading from right to left, the census enumerations are numbered as 1, ..., $N_L$, where $N_c$ of them can be linked to $\mc{P}$. The linked part of $\mc{P}$ is referred to as the \emph{core} of PD, denoted by $\mc{P}_c$. One observes $\delta(k\in U) = 1$ and $\delta(\mbox{adr}_k = \bs{a}_k)$ for all $k\in \mc{P}_c$, based on the census data. One can treat $\mc{P}_c$ as a non-probability sample from $\mc{P}$. Under the assumption of non-informative selection, i.e.
\[
\bs{\mu}(\bs{a}_k, \bs{z}_k | k\in \mc{P}_c) = \bs{\mu}(\bs{a}_k, \bs{z}_k | k\in \mc{P} \cap U) ~,
\]
one can estimate $\bs{\mu}(\bs{a}_k, \bs{z}_k | k\in \mc{P} \cap U)$ consistently from $\mc{P}_c$. This allows one to populate the probabilities $\xi_k$ and $\bs{\mu}_k$ for all $k\in \mc{P}$, in the case of $k\in U$. Moreover, these probabilities can be benchmarked to the census population estimates, denoted by $\widehat{N}$, at any aggregation level or for any sub-population of choice. There are different ways of benchmarking; see e.g. Favre et al (2005) for a method that can incorporate unit-level constraints. 

The estimation of the probability of erroneous enumeration, denoted by $\theta_k$ for $k\in \mc{P}$, requires a different approach. Some possibilities are given below.
\begin{itemize}[leftmargin=5mm]
\item One can replicate the approach of Latvia and Estonia, whereby one labels a subset of $\widehat{N}$ persons in $\mc{P}$, denoted by $\mc{P}_U \subset \mc{P}$. The additional non-core persons in $\mc{P}_U \setminus \mc{P}_c$ are the ones judged to be the most likely in-scope persons in $\mc{P}\setminus \mc{P}_c$ or, equivalently, the persons in $\mc{P}\setminus \mc{P}_U$ are the most likely out-of-scope persons in $\mc{P}\setminus \mc{P}_c$. 

\item One can use the population hypercube estimated from the census and census coverage survey, and assign the probabilities $\theta_k$ according to the cell of person $k$, for $k\in \mc{P}$.   

\item One can draw a probability sample $s$ from $\mc{P}\setminus \mc{P}_c$ and obtain $\delta(k\in U)$ for $k\in s$, and use the combined sample $\mc{P}_c \cup s$ to estimate $\theta_k$, for $k\in \mc{P}$. 
\end{itemize}
Benchmarking of the estimated $\theta_k$'s may be necessary, denoted by $\widehat{N} + \sum \theta_k = N_p$, in the case of individual-based estimation. Finally, notice that international migration and other special populations may need to be treated separately from the above.

\subsection{Basic ideas of rolling}

By \emph{rolling} we mean that in principle the fractional counters can be updated in a nearly continuous manner over time, just as $\mc{P}$ itself. It seems most similar to \emph{incremental learning} in the statistical machine learning literature. Below we first summarise the data that can be made available for rolling, and then discuss the basic ideas in the parametric and algorithmic settings, respectively.

\subsubsection{Data for rolling}

Let $\mc{P}_t$ be the PD at time $t$, where $t\geq 0$, and $\mc{P}_0$ is the initiation PD, say, at the census time point. Denote by $\mc{L}_t$ the set of labelled persons, where $\mc{L}_t \subseteq \mc{P}_t$, on which supervised learning can be based. Without losing generality, one can partition $\mc{L}_t$ into three parts:
\begin{equation} \label{data}
\mc{L}_t = S_t \cup \mc{B}_t \cup \mc{A}_t
\end{equation}
where $S_t$ denotes the subset of persons that are associated with known inclusion probabilities, and $\mc{B}_t$ denotes the other persons for whom we observe updated labels of (erroneous, displaced, placement), and $\mc{A}_t$ the rest for whom we have only the labels from $t-1$. \textit{It will be important and helpful to enhance the collection, organisation and usage of all the relevant data, across the ONS, in order to facilitate efficient rolling and enable the transition to a sustainable system for population statistics in the long term.}
 
\paragraph{Coverage survey} The planned post-census coverage survey is a source of $S_t$. The observations in $S_t$ can obviously be used for updating of $\bs{\mu}$ and $\xi$, pertaining to the probabilities of displaced and placement. Depending on the actual design and estimation method, it may as well be possible to update the erroneous enumeration probability $\theta$, especially if it is possible to administer follow-up surveys at the sampled addresses in the coverage survey.
 
\paragraph{On-going surveys} As mentioned before, the planned approach at Istat is to draw household samples for the main social surveys from their first-phase sample for population statistics. Unless the ONS adopts the same approach, there will be other labelled persons from the on-going social surveys, in addition to the coverage survey. Notice that the fieldwork protocol in the on-going surveys will need to be enhanced to ensure the quality of the data collected for rolling. Whether these additional labelled persons can be classified as part of $S_t$ or $\mc{B}_t$ depends on the actual sampling design. 

\paragraph{Administrative sources} Updating in the relevant administrative sources can generate labelled persons. For instance, updating of the Council Tax register, the home addresses in Drivers' License database, the PAYE register, and so on, can all provide data for $\mc{B}_t$. The distinction of core and non-core in $\mc{B}_t$ can be relevant at least in the near future.

\subsubsection{Rolling in the parametric setting} \label{EBP}

For a parametric setting of the fraction counters, suppose the relevant probabilities are given by the inverse of the logistic link function, denoted by
\[
\bs{\pi}(\bs{x}_k, \bs{\beta}) = E(\bs{y}_k | \bs{x}_k ; \bs{\beta})
\]
for person $k$, where $\bs{y}_k$ is the vector of indicators whose components sum to 1, and $\bs{x}_k$ is the vector of known covariates, and $\bs{\beta}$ the logistic regression coefficients. Suppose the initial $\bs{\beta}_0$ are estimated based on a large dataset in connection with the census, denoted by $\hat{\bs{\beta}}_0$, with associated variance $\hat{\bs{\Sigma}}_0$. 

At the next time point $t=1$, one could refit the model using the labelled persons in $\mc{L}_1$, of which $\mc{D}_1 = S_1\cup \mc{B}_1$ are associated with updated observations of $\bs{y}_{1,k}$, for $k\in \mc{D}_1$. This assumes $\bs{y}_k$ remains the same at $t=0, 1$, for any non-updated person $k\in \mc{L}_1 \setminus \mc{D}_1$, which may be problematic if the lack of updating is due to delays or errors in the sources. One could refit the model using only the data associated with $\mc{D}_1$, under some suitable assumption of non-informative selection of $\mc{D}_1$ from $\mc{P}_1$. The estimation precision is then determined by the size of $\mc{D}_1$, which is much smaller than the initiation dataset $\mc{L}_0$, so that the uncertainty of the estimated $\bs{\beta}_1$ will be much larger than that of $\hat{\bs{\beta}}_0$.

Consider empirical Bayes (best) prediction (EBP) under the hierarchical model:
\begin{gather*}
 E(\bs{y}_{1,k} | \bs{x}_{1,k} , \bs{\beta}_1) = \bs{\pi}(\bs{x}_{1,k}, \bs{\beta}_1) \\
\bs{\beta}_1 \sim N(\hat{\bs{\beta}}_0, \hat{\bs{\Sigma}}_0)
\end{gather*}
where the normal distribution is motivated by the large size of initiation dataset $\mc{L}_0$. At the lower level, the population dynamics which change the parameter $\bs{\beta}_0$ from time 0 to $\bs{\beta}_1$ at time 1 is modelled as a `random'  departure from the previous `position' $\hat{\bs{\beta}}_0$ with variance $\hat{\bs{\Sigma}}_0$. This differs in concept to fully Bayesian approach, where the hyper-parameter of the prior distribution of $\bs{\beta}_1$ needs not to have any empirical connotation.   

Assuming IID observations over $\mc{D}_1$, we obtain the prediction function for $\bs{\beta}_1$ as
\[
f(\bs{\beta}_1 | \bs{y}_{1,\mc{D}_1}, \bs{x}_{1,\mc{D}_1} ; \hat{\bs{\beta}}_0, \hat{\bs{\Sigma}}_0) = 
\frac{\prod_{k\in \mc{D}_1} f(\bs{y}_{1,k} | \bs{x}_{1,k}, \bs{\beta}_1) \cdot 
\phi(\bs{\beta}_1; \hat{\bs{\beta}}_0, \hat{\bs{\Sigma}}_0)}{\prod_{k\in \mc{D}_1} f(\bs{y}_{1,k} | \bs{x}_{1,k})}
\]
Let $\hat{\bs{\beta}}_1$ and $\hat{\bs{\Sigma}}_1$ be the prediction mean and variance of $\bs{\beta}_1$, respectively. In this way the lower-level model is updated to $\bs{\beta}_2 \sim N(\hat{\bs{\beta}}_1, \hat{\bs{\Sigma}}_1)$, by which the model is rolled forward and ready to be used for updating at $t=2$.

The EBP approach can thus facilitate the rolling of fractional counters, without the extra and potentially problematic assumption that $\bs{y}_k$ remains the same for $k\in \mc{L}_1 \setminus \mc{D}_1$, or losing efficiency as when estimating $\bs{\beta}_1$ only based on $\mc{D}_1$. It achieves stability over time, balancing between the signals from $\mc{D}_t$ and the inertia of $N(\hat{\bs{\beta}}_{t-1}, \hat{\bs{\Sigma}}_{t-1})$: any value of $\bs{\beta}_t$ far from $\hat{\bs{\beta}}_{t-1}$ are `weighted down' by $\phi$, compared to only based on $f(\bs{y}_{t,k} | \bs{x}_{t,k}, \bs{\beta}_t)$.

\subsubsection{Rolling in the algorithmic setting} \label{learning}

Machine learning or statistical machine learning has a vast and rapidly growing literature. There does not exist a unified framework of the different approaches. Below we first list some classifications and concepts that seem relevant, before we illustrate and discuss the basic ideas of decision tree updating in the present context.

With respect to the logical basis of learning, a distinction is between transduction and induction. Transduction or transductive inference is reasoning from observed (training) cases to specific (test) cases. In contrast, induction is reasoning from observed training cases to general rules, which are then applied to the test cases. The distinction seems most interesting in siutations where the predictions of the transductive model are not achievable by any inductive model. However, transductive algorithms that seek to predict discrete labels tend to be derived by adding partial supervision to a clustering algorithm, while supervised learning is generally considered to be a form of induction. 

With respect to the context of learning, one commonly distinguishes among unsupervised learning (without labelled units), supervised learning (based on labelled units), and reinforcement learning (in an interactive environment). The last type of algorithm enables an agent to learn by trial and error using feedbacks from its own actions and experiences, such as in gaming. Semi-supervised learning is a class of techniques that typically use a small amount of labelled data with a large amount of unlabelled data, as many researchers have found that unlabelled data, when used in conjunction with a small amount of labelled data, can produce considerable improvement in learning accuracy. 

With respect the process of learning, broadly speaking there are two ways to train a model. A static model is trained offline, once and then used as-is for a while. A dynamic model is trained online, where data is continually entering the system and incorporated into the model through frequent updates. 
\begin{itemize}[leftmargin=4mm]
\item In active learning, one seeks to select the most informative unlabelled instances and ask an omniscient oracle for their labels, in order to retrain a learning algorithm to maximise accuracy. Clearly, the selection mechanism can be designed to resemble audit sampling for model validation or improvement. 

\item In incremental learning, the data is generated by an external process, and continually used to further train the model, i.e. without the model being completely retrained using all the available data at any given time point. The motivation may be technical, e.g. the data becomes available only gradually over time or its size is out of system memory limits. A central concern is to allow the model to adapt to new data without forgetting its existing knowledge. Some incremental learners have built-in parameters or assumptions that control the relevancy of new and old data. See e.g. Ade and Deshmukh (2013), Gepperth and Hammer (2016).
\end{itemize}

Regarding the validity of the trained model, data or concept shift is a term one finds in the machine learning literature, which is said to occur when the joint distribution of inputs and outputs differs between training and test stages. Covariate shift, a particular case of data shift, occurs when only the input distribution changes. An example is email spam filtering, which may fail to include spams (for training) that are not detected by the filter used to classify spams. Relevant statistical concepts developed for various informative selection mechanisms do not seem to have attracted much attention here.

\subsubsection{Rolling of a decision tree} \label{tree}

The Very Fast Decision Tree (VFDT) system is one of the most successful algorithms for mining data streams (Domingos and Hulten, 2000). Its main innovation is the use of Hoeffding bound to decide how many examples (observations) are necessary to observe before installing a split-test at a leaf of the tree. Splitting the leaf makes only a local change to the tree, since the prediction of units ending at another leaf is not affected. The theoretical result refers to the maximum expected disagreement between the Hoeffding tree and the asymptotic batch tree given infinite observations. However, the asymptotic batch tree is hardly our present interest, where one may assume that the population structure (hence the tree itself) must change over time, and the target is not the tree that one will grow given infinite observations that span infinitely over time. 

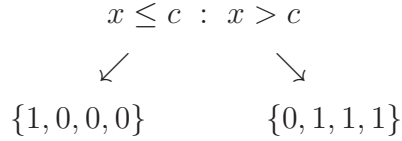
\begin{align*}
x\leq c ~ & : ~ x> c \\
\swarrow \hspace{8mm} & \hspace{10mm} \searrow \\
\{ 1, 0, 0, 0\} \hspace{6mm} & \hspace{10mm} \{ 0, 1, 1, 1\}
\end{align*}

To illustrate the issue, consider the above split for two leaves in the tree grown at $t-1$. Let there be one observation $(x', y')$ at $t$ passing this way. Now,
\begin{itemize}[leftmargin=6mm]
\item if $x' \leq c$ and $y' = 1$, or $x' > c$ and $y' = 0$, then $(x', y')$ may be considered to provide `negative' evidence to the current tree, but perhaps not necessarily so if the value 1 in the left leaf or 0 in the right leaf happens to be observed a long time ago;

\item if $x' \leq c$ and $y' = 0$, or $x' > c$ and $y' = 1$, then $(x', y')$ may be considered to provide `positive' evidence to the current tree, but perhaps not necessarily so if a value 0 in the left leaf or 1 in the right leaf happens to be observed a long time ago.
\end{itemize}
In other words, for the rolling of a decision tree that evolves over time (subjected to concept shift), more considerations are required than the number of observations and the discriminant measure of the split.  

It seems that one may still need to use part of the updated observations in $\mc{D}_t$ for training and part of them for validation. Let $M_t$ denote the updated tree, and $M_{t-1}$ the tree grown at $t-1$. At least two measures may be needed: 
\begin{itemize}[leftmargin=16mm]
\item[$\Delta_{\epsilon}$:] how much better $M_t$ predicts for the updated units in $\mc{D}_t$ than $M_{t-1}$,

\item[$\Delta_M$:] how much change $M_t$ predicts for the non-updated units in $\mc{P}_t \setminus \mc{D}_t$.
\end{itemize}
Since $M_{t-1}$ yields 0 in terms of both measures, one may need to balance between the two measures when growing $M_t$. For instance, one may choose to maximise $\Delta_{\epsilon}$ subjected to an upper bound on $\Delta_M$, or minimise $\Delta_M$ subjected to a lower bound on $\Delta_{\epsilon}$.

\section{Register-based auditing-assisted approach} \label{RBAA}

Greater resource savings will be the case in future, if population statistics are produced from administrative data and on-going surveys under the APD approach, while purposely designed coverage survey is only used from time to time for \emph{auditing}. There are two necessary elements to such a  \textit{register-based auditing-assisted} APD approach.

\paragraph{Rolling without coverage surveys} At some stage one needs to be able to exclude the envisaged regular coverage survey data from $\mc{L}_t$ in \eqref{data}, and only use the updated observations from other on-going surveys and administrative sources for the rolling of fractional counters, whether it is in the parametric or algorithmic setting. 

This is not as unthinkable as it may seem at first. For instance, it may be noticed that the existing DBE approach is essentially a register-based APD approach, based on an estimated population hypercube. The Irish APD approach suggests it may be possible to estimate the population hypercube purely based on administrative data, albeit using a different methodology than DBE component estimation. In the register-based APD approaches of Israel, Latvia and Estonia, individual counters are produced by different methods, none of which uses any regular coverage surveys. The Norwegian household register provides another example, where decision rules are applied to individual-level data from the administrative sources. So the question that matters is how good a register-based individual-level APD approach can become in the UK. 

\paragraph{Auditing inference} Zhang (2018) contrasts register-based APD approach to the APD approach based on combining registers and coverage survey. Under the purely register-based approach, ``an estimator, denoted by $\hat{N}_n$, can be calculated under a statistical model, using multiple incomplete administrative registers, where $N$ denotes the unknown population size and $n$ the generic size of the available datasets.'' It is suggested that, regardless how effective the rolling of $\hat{N}_n$ may be, one is unlikely ``to have $\hat{N}_n / N \stackrel{P}{\rightarrow} 1$ under some asymptotic setting, as $n, N\rightarrow \infty$''. Audit sampling will be necessary, in order to ``validate the model underlying $\hat{N}_n$, ..., which is affected by the sampling error of the auditing survey''. This raises the challenge of audit sampling inference. 

Using disaggregation of Consumer Price Index based on proxy household expenditure measures obtained from transaction data, Zhang (2019a) develops an audit sampling inference approach for big-data statistics. Generically speaking, let $\theta_0$ be the true scalar parameter value of interest. Let $\theta^*$ be a point estimate based on big data, such that its variance is negligible compared to its potential bias for all practical purposes. One can test $H_0: \theta^* = \theta_0$ based on audit sampling. However, an accuracy measure is needed, even if the hull hypothesis cannot be rejected at the chosen level. A general dilemma in this context is the following. Let $\hat{\theta}$ be an unbiased estimator of $\theta_0$ based on audit sampling, and let $\widehat{V}(\hat{\theta})$ be an unbiased estimator of its audit sampling variance. An unbiased estimator of the mean squared error (MSE) of $\theta^*$ can be given as
\[
\widehat{\mbox{MSE}} = (\theta^* - \hat{\theta})^2 - \widehat{V}(\hat{\theta}) ~.
\] 
However, when the bias $\theta^* - \theta$ is small, auditing may fail to yield a meaningful measure, if the audit sampling variance is not small enough, in which case one easily obtains $\widehat{\mbox{MSE}} < 0$ as the result. To overcome the dilemma, Zhang (2019a) proposes a novel accuracy measure to replace the standard MSE. If feasible in the present context, then one can employ an audit sample that is much smaller than the envisaged coverage survey sample.

\paragraph{Summary} In the long-term perspective, greatest gains can be achieved via a gradual transition from an APD approach that requires coverage survey sampling to a register-based auditing-assisted APD approach. Audit sampling aims to validate the register-based statistics, and to generate meaningful accuracy measure, for which one can use a much smaller sample size than the envisaged coverage survey sample. To enable the transition, it will be important that one as soon as possible starts the development, so that one can test and refine the required methods and to obtain the necessary experience and confidence over time.

\section{Key topics for methodology development} 

Four inter-related topics for methodological development emerge from the above: 
\begin{itemize} \setstretch{1.0}
\item data linkage for longitudinal PD;
\item rolling or incremental learning, including benchmarking, design of coverage survey, and parallel development of register-based auditing-assisted APD approach;
\item appropriate uncertainty propagation or accuracy measure in various scenarios;
\item methodology for producing social statistics in the new environment. 
\end{itemize}

\subsection{Longitudinal PD} 

\textit{Generic scalable linkage methodology is a premise to the provision of UK neighbourhood population statistics}. 
In the first instance, the relevant longitudinal administrative data should be linked to form the longitudinal PD; in the next instance, longitudinal data linkage is the ability to link the longitudinal PD to other open or free data, as well as relevant sample surveys (often longitudinal themselves). 

Davis-Kean et al. (2017) projects an ambitious outlook to the longitudinal population for ESRC UK longitudinal study resources. In particular, this aims at standardising the designs of the various longitudinal surveys so that they all use the same \emph{longitudinal population register} (i.e. a population spine) as their sampling frame, and with all ESRC research-related linkage of different administrative and survey data sources harmonised to this spine. As an example of such a constructed longitudinal population spine, in countries that do not have a population register to start with, one may refer to the Integrated Data Infrastructure (IDI) at Statistics New Zealand (SNZ, 2018). 

Although the Fellegi and Sunter (FS) methodology for record linkage has proven to be very useful in practice (e.g. Owen et al., 2015), it does have some theoretical issues.
\begin{itemize}[leftmargin=6mm]
\item Applying the Likelihood Ratio Test to all the pairs $A\times B$ in files $A$ and $B$ creates a multiple comparison problem. The acceptable pairs require deduplication, e.g. when both $(ab)$ and $(ab')$ are above the acceptance threshold. It is difficult to link multiple files in a transitive manner, e.g. that $(ab)$ in $A\times B$ and $(bc)$ in $B\times C$ are links does not necessarily entail $(ac)$ will be accepted as a link when looking at $A\times C$.
 
\item The joint distribution of all the $n_A n_B$ comparison scores is ill-defined, if one treats e.g. the comparison scores for $(ab)$ and $(ab')$ as if they were independent of each other. The so-called maximum likelihood estimator of the parameters of the $m$- and $u$-probabilities (Jaro, 1989) are biased in reality; see e.g. Fortini and Tuoto (2019).
\end{itemize}

Entity resolution provides theoretically a more attractive formulation (e.g. Christen, 2012), where the set of (unique) entities underlying the separate datasets are envisaged as a latent spine of unknown size, and each record (in any dataset) is attached to one and only one latent entity on the spine. In this way, the records in different files are linked to each other or deduplicated, provided they are attached to the same latent entity, in a transitive manner regardless of the number of datasets involved. There are a few applications of the entity resolution perspective under the Bayesian paradigm of computation (e.g. Tancredi and Liseo, 2011; Stoerts et al., 2017), although there is nothing intrinsically Bayesian about this perspective to record linkage. Lack of scalability has been a central challenge to the proposed methodology so far, which is not yet feasible e.g. to link the population census file with the patient register.

Scalable linkage methods for multiple population-size datasets are important to the creation of the longitudinal PD. Moreover, the generic ability to link multiple files in a transitive manner can be expected to improve the quality of statistical information harnessed in the linked dataset. Replacing the FS-paradigm to record linkage by the entity resolution perspective can provide the angle for innovative approaches.

\subsection{Rolling or incremental learning}

\textit{The methodology of rolling or incremental learning needs to be studied and decided upon.}
Firstly, one needs to find out how rolling by EBP in the parametric setting (Section \ref{EBP}) works out in practice. Next, it is possible that algorithmic learning (Section \ref{learning}) can provide a more flexible and powerful predictive modelling approach. However, incremental learning in the presence of concept shift (i.e. the model changes over time) does not yet have an established approach in the literature. Methodological developments in this respect will be necessary, in case one would like to pursue algorithmic learning. 
 
Whether one adopts parametric or algorithmic learning, there are at least three other relevant aspects that seem worth attention, as discussed below.

Firstly, one may naturally wish to benchmark the updated fractional counters, towards the estimated population hypercube from combined register and coverage survey data, as during their initiation (Table \ref{tab-initial}). A possibility is to only apply benchmarking when producing population statistics to be disseminated, say, on a yearly basis in the years immediate after 2021. In other words, benchmarking and rolling of fractional counters can have different frequencies. The methodology and practice need to be established.

Secondly, how to make effective use of the coverage survey, and can active learning be related to its design? By active learning, one seeks to observe the unlabelled instances that are most effective for model validation or improvement. For instance, in the present context, it seems reasonable that one should give higher sample representation of the non-core part of the PD, or the persons who are judged to have weak fractional counters, e.g. placement probabilities $\bs{\mu}_k$ are not `close' to a dummy vector, or a relatively large probability of being displaced ($\xi_k$) or out-of-scope ($\theta_k$). Regardless of the exact characterisation, this suggests that one may need a method for follow-up surveys of some addresses, given the PD-status of the persons at different addresses.

Thirdly, `parallel' learning of register-based fractional counters (Section \ref{RBAA}) requires a different approach, where the coverage survey data are only used indirectly for updating of the model, but not directly as labelled observations for supervised learning. This means that one would like to combine supervised learning from $\mc{L}_t \setminus \mc{C}_t$, where $\mc{C}_t$ denotes the coverage survey sample, and heuristic updating of the fitted model, based on the evidence in $\mc{C}_t$. Such heuristic updating differs from benchmarking, where the latter requires estimates using the coverage survey. For instance, one may envisage heuristic updating in the form of a decision tree, where the paths close to the root are determined by stable decision rules evidenced from the comprehensive coverage survey data, while the observations in $\mc{L}_t \setminus \mc{C}_t$ are only used for the splits and sprouting near the leaves.

\subsection{Uncertainty propagation or accuracy measure}

\textit{Theoretical conceptualisation and practical method for uncertainty propagation or accuracy measures are needed in several scenarios}. 

Firstly, for the initiation of fractional counters, the core part of PD that can be linked to the census enumerations will be used for the estimation of displacement and misplacement probabilities, whereas the census enumerations and the census coverage survey will be used for estimating the probability of erroneous enumeration. It needs to be verified whether conditional uncertainty propagation given the estimated fractional counters can sensibly capture the various underlying variations, and if not, how it might be modified to produce plausible accuracy measures in a practical manner. 

As explained in Section \ref{EBP}, uncertainty propagation from parameter updating to fractional counting given the parameter can be based on a coherent scheme under the rolling of parametric EBP. The matter is less clear under incremental algorithmic learning. Suppose the fitted model is given as a decision tree, which is updated by a constrained optimisation method (Section \ref{tree}). Conditional uncertainty propagation given the updated tree is straightforward, just as when the fractional counters are given parametrically. But how can one incorporate the uncertainty of the tree updating itself? One can introduce some kind of bootstrap. But at which level should one allow the replicate trees to vary from each other: simply where the new splits are created, or higher up? 

Finally, the methodology of audit sampling inference needs to be worked out for register-based fractional counting (Section \ref{RBAA}). It will be possible to treat the coverage survey sample, or part of it, as the audit sample, whether the coverage survey is designed to accommodate active learning or not. This is necessary in order to provide the statistical argument for the transition towards a register-based auditing-assisted APD approach in the long term, in replacement of the more costly continuous coverage survey.

\subsection{Producing social statistics in the new environment}

Though not central to this report, it must be pointed out that 
\textit{the production of social statistics requires a broad perspective to design and estimation in the new environment}. 

The specifics of future provision of population statistics will change considerably in the UK. Traditional MYEs with decennial census updating will no longer be the foundation of social statistics on temporally varying topics and phenomena of interest. It would be narrow-minded and ineffective to simply replace the population benchmarks, albeit produced based on a different APD approach beyond 2021, but keep the same design strategies across the spectrum of social survey programmes.

For instance, as mentioned earlier, a two-phase approach is currently being developed in Italy, where the first-phase sample targets mainly at the population statistics, and the major social survey samples are selected as negatively coordinated second-phase samples. It is yet unclear whether this is a suitable solution in the UK. Neither is it necessarily the most effective approach, generally speaking or in the long run, when it comes to the combined use of coverage survey and other social social surveys. 

The coverage survey as currently envisaged may no longer be necessary, provided the transition to a register-based auditing-assisted APD approach to population statistics. How to combine audit sampling and on-going social surveys will be then a different overall design question. Indeed, greater use of administrative data is expected to extend to the area of social statistics as well. It is thus perhaps appropriate to set on a future landscape of register-based auditing-assisted population \emph{and} social statistics.

\end{document}